\documentclass[12pt,preprint]{aastex}
\def\jcap{JCAP}
\def\beq{\begin{equation}}
\def\eeq{\end{equation}}
\def\ben{\begin{eqnarray}}
\def\een{\end{eqnarray}}
\def\hbs{\hat{\bf S}}
\def\hbp{\hat{\bf p}}

\def\bs{{\bf S}}
\def\bx{{\bf x}}

\def\delx{\delta ({\bf x})}
\def\delk{\tilde{\delta}({\bf k})}
\def\lam{\lambda}
\def\mt{\Delta m_{\rm t}}
\def\munit{\,h^{-1}\! M_{\odot}}
\def\dunit{\,h^{-1}{\rm Mpc}}
\def\lm{\log [M_{h}/(h^{-1}M_{\odot})]}
\def\cth{\cos\theta}
\def\cthi{\cos\theta_{i}}

\def\mcthi{\langle\cos\theta_{i}\rangle}

\def\mctha{\langle\cos\theta_{1}\rangle}
\def\mcthb{\langle\cos\theta_{2}\rangle}
\def\mcthc{\langle\cos\theta_{3}\rangle}
\begin{document}
\title{The Halo Spin Transition as a Probe of Dark Energy}
\author{Jounghun Lee\altaffilmark{1}, Noam I Libeskind\altaffilmark{2,3}}
\altaffiltext{1}{Department of Physics and Astronomy, Seoul National University, Seoul 08826, Republic of Korea  
\email{jounghun@astro.snu.ac.kr}}
\altaffiltext{2}{Leibniz-Institut f\"{u}r Astrophysik Potsdam (AIP), An der Sternwarte 16, 14482 Potsdam, Germany 
\email{noam@aip.cn.kr}}
\altaffiltext{3}{University of Lyon, UCB Lyon-1/CNRS/IN2P3, IPN Lyon, France}

\begin{abstract}
We present a numerical evidence supporting the claim that the mass-dependent transitions of the halo spin orientations from the intermediate 
to the minor principal directions of the local tidal fields can in principle be a useful discriminator of dark energy models. 
We first define a {\it spin transition zone} as the mass range of the halos, $\mt$, for which the intrinsic spin alignments with the minor tidal principal 
directions become as strong as that with the intermediate principal directions.  Then, utilizing the halo samples from the DEUS simulations 
performed separately for the WMAP7 $\Lambda$CDM, phantom DE and quintessence models, we investigate if and how the three different dark energy 
models differ in $\mt$. 
It is shown that the differences in $\mt$ among the three dark energy models are significant enough to discriminate the models from one 
another and robust against the variations of the smoothing scale of the tidal field and redshift.  Noting that a narrower spin transition zone is more 
powerful as a probe of dark energy,  we also show that the spin transition zones become narrower at higher redshifts, in the filamentary 
environments and for the case that the tidal fields are smoothed on the smaller scales.   
Our result is consistent with the scenario that $\mt$ is mainly determined by how fast the nonlinear evolution of the tidal field proceeds, which in turn 
sensitively depends on the background cosmology.
\end{abstract}
\section{Introduction}\label{sec:intro}

Ever since the best-fit theoretical curves to the observed luminosity-distance relations of the type Ia supernovae (SN Ia) were found to be  
obtained from the Friedmann equations with non-vanishing cosmological constant term $\Lambda$ \citep{rie-etal98,per-etal99},  it has been widely 
accepted that the present universe is in an accelerating phase caused by the negative pressure of the dominant energy content $\Lambda$ with 
equation of state $w=P_{\Lambda}/\rho_{\Lambda}=-1$.  
Backed up further by the observations of the cosmic microwave background (CMB) temperature power spectrum and the large scale structures 
\citep[e.g.,][and references therein]{teg-etal04,wmap7}, 
what has been established as a standard model of cosmology for the past two decades is an inflationary flat universe containing the dominant 
$\Lambda$ and non-baryonic cold dark matter (CDM), being often dubbed the $\Lambda$CDM cosmology. 

The solidity of the $\Lambda$CDM cosmology as a physical model has been known to be overcast by two conceptual problems associated 
with $\Lambda$. First, its value inferred from the SN Ia observation is too low to be explicable by any fundamental physics. 
Second, it requires unnaturally fine-tuned initial conditions to explain the observed ratio of the matter to $\Lambda$ density parameter being an 
order of unity, $\Omega_{m}/\Omega_{\Lambda}\sim 1$, at the present epoch \citep{car01,car06}. 
Nonetheless, the simplicity and effectiveness of the $\Lambda$CDM cosmology in explaining the large-scale features of the universe at a quantitative 
level has justified connivance of the community at these profound problems \citep{spr-etal06}.

The latest observational data measured with unprecedentedly high precision and analyzed with the cutting-edge statistical techniques, however, 
have signaled a rupture between the early and the late universes described by the $\Lambda$CDM model. The most notorious example of this rupture is the 
value of the Hubble constant, $H_{0}$, obtained from the observations of the local Cepheids, which turned out to be in $4.4\sigma$ 
tension with its best-fit value from the latest CMB analysis \citep{rie-etal19,planck18}. 
Another example is the amplitude of the linear density power spectrum, $\sigma_{8}$ and density parameter, $\Omega_{m}$, whose best-fit values 
from the cosmic shear surveys showed mismatches with the central values from the latest CMB analysis under the assumption of a $\Lambda$CDM 
universe \citep{abb-etal18,hik-etal19}.  
Although various attempts have been made to reconcile the near-field measurements with the CMB best-fits 
\citep[e.g.,][]{agr-etal19,pou-etal19,kre-etal20}, none of them have so far been fully satisfactory.

Being progressively perceived as a cardinal drawback of the $\Lambda$CDM model rather than being casually ascribed to some unidentified 
systematics, the aforementioned rupture brought out the importance and necessity of developing competitive near-field probes to distinguish between 
the viable alternative DE and the $\Lambda$CDM models, which equally satisfy the CMB constraints. What has been sought after as an optimal 
near-field probe is the one that describes the small-scale features of the universe in the deeply nonlinear regime and concurrently possesses a sensitive 
dependence on the background cosmology.

Very recently, \citet{lee-etal20} suggested a new near-field diagnostics based on the mass dependent transition of the halo spins, which refers to the 
numerically and observationally detected phenomenon that while the high-mass halos have their spins preferentially aligned with the directions perpendicular 
to the minor principal axes of the local tidal fields, the low-mass ones prefer the directions parallel to them in their spin orientations 
\citep[e.g.,][]{ara-etal07,hah-etal07b,paz-etal08,cod-etal12,TL13,tem-etal13,tro-etal13,lib-etal13,AY14,dub-etal14,for-etal14,cod-etal15a,cod-etal15b,hir-etal17,
cod-etal18,gan-etal18,wan-etal18,gan-etal19,lee19,kra-etal20}.  What \citet{lee-etal20} newly found was that the critical mass at which the transition of the halo 
spin orientations occurs (transition mass of the halo spins) varies sensitively with the total neutrino mass, regardless of the smoothing scale of the tidal field. 

In light of \citet{lee-etal20}, we set off here to numerically explore if the transition mass of the halo spins can be used as a probe of DE with the 
help of $N$-body simulations. The contents of this Paper can be outlined as follows. In Section \ref{sec:data} we will describe a dataset from N-body 
simulations for three different dark energy models and present a routine with which the mass dependent transitions of the halo spins are determined. 
In Section \ref{sec:zone}, we will present a numerical evidence for the usefulness of the halo spin transitions as a discriminator of dark energy models. 
In Section \ref{sec:web}, we will describe how the sensitivity of the halo spin transitions as a probe of dark energy varies with the cosmic web type. 
In Section \ref{sec:con}, we will summarize the main results and discuss their implications. 

\section{Effect of DE on the Transition Mass of the Halo Spins}
\subsection{Data and Routine}\label{sec:data}

For the current study, we consider three different DE models, namely, a flat $\Lambda$CDM model with equation of state $w=-1$, a phantom DE model with 
negative constant equation of state $w<-1$, and a quintessence model with time varying equation of state $-1<w(t)<-2/3$,  simulated by the Dark Energy 
Universe Simulation (DEUS) project in a periodic box of comoving volume $648^{3}\,h^{-3}$Mpc$^{3}$ with $2048^{3}$ DM particles 
\citep{deus1,deus2,DEUS}.  
For the flat $\Lambda$CDM model, the six key cosmological parameters were set at the central values determined by 
the Wilkinson Microwave Anisotropy Probe Seven Year data (WMAP7) \citep{spe-etal07,wmap7}.  
For the other two DE models, the values of the seven cosmological parameters (the six key parameters + $w$) were chosen to satisfy 
the same WMAP7 constraints. Table \ref{tab:initial} lists the values of the matter density parameter, $\Omega_{m}$, linear power spectrum amplitude on 
the scale of $8\dunit$, $\sigma_{8}$,  dimensionless Hubble parameter, $h$, dark energy equation of state, $w\equiv w_{0}+w_{a}(1-a)$ with scale factor 
$a(t)$ \citep{lin03}, and mass of individual DM particles, $m_{p}$, for the three DE models. For more detailed description of the DEUS, we refer the readers 
to \citet{deus1} as well as the DEUS webpage 
\footnote{All data from the DEUS are publicly available at http://www.deus-consortium.org/}. 

A phantom DE ($w$CDM) is characterized by its negative kinetic energy that exerts stronger repulsive force than $\Lambda$ \citep{phantom}. 
Whereas, a quintessence model is characterized by the potential shape of a scalar field DE. For the DEUS was specifically chosen the Ratra-Peebles 
potential whose shape is expressed as powers of linear combinations of the exponentials of the scalar field \citep{RP88}. 
Although the distant-field diagnostics based on the linear observables like the CMB temperature spectra and SN Ia luminosity-distance relations cannot 
distinguish among the three models \citep{DEUS}, an efficient near-field probe based on the nonlinear observables might be able to pull it off since the formation 
and evolution of the nonlinear structures proceed in a different way among them \citep[e.g., see][]{ali-etal10}.
 
The halo catalog from the DEUS comprises the gravitationally bound objects identified at various redshifts by the Friends-of-Friends (FoF) halo finder with 
a linkage parameter of $0.2$ \citep{roy-etal14} and provides information not only on the center of mass position, $\bx$, and virial mass, $M_{h}$, of each 
object but also on the positions and velocities of its constituent DM particles, which allow us to compute its angular momentum vector, $\bs\equiv S\hbs$. 
The well-resolved FoF objects in the mass range of $11.8\le \log [M_{h}/(\munit)]\le 13$ at $z=0$ are selected for a sample of the galactic halos to be 
used for the current analysis. The low-mass FoF objects with $\log [M_{h}/(\munit)] < 11.8$ are excluded from our analysis on the ground that the angular momentum 
vectors of the halos composed of less than $300$ DM particles are likely to suffer from inaccuracy caused by the shot-noise \citep{bet-etal07}.

The routine we go through for the investigation of the DE dependence of the halo spin transitions is similar to that described in \citet{lee-etal20}, 
the summary of which is provided in the following.
\begin{itemize}
\item
Divide the simulation volume into $256^{3}$ grids and create a density field, $\rho({\bf x})$, on the grids by applying the cloud-in-cell method 
to the DM halo distributions from the DEUS. Calculate the dimensionless density contrast field, 
$\delx\equiv \left[\rho({\bf x})-\langle\rho\rangle\right]/\langle\rho\rangle$ where $\langle\rho\rangle$ 
is the spatial average of $\rho({\bf x})$ taken over the $256^{3}$ grids.
\item
Evaluate the Fourier-space density contrast field, $\delk$, by applying the Fast-Fourier Transformation (FFT) method to $\delx$, where 
${\bf k}=(k\hat{k}_{i})$ is the Fourier space wave vector.
Create a tidal shear field, ${\bf T}({\bf x})=\left[T_{ij}({\bf x})\right]$, smoothed with a Gaussian filter on the scale of $R_{f}$ via the inverse 
Fourier transformation of $\delk\exp(-k^{2}R^{2}_{f}/2)\hat{k}_{i}\hat{k}_{j}$. At each grid, diagonalize 
$T_{ij}({\bf x})$ to determine its three eigenvalues, $\{\lam_{1},\ \lam_{2},\ \lam_{3}\}$ with $\lam_{1}\ge\lam_{2}\ge\lam_{3}$, and corresponding 
orthonormal eigenvectors, $\{\hbp_{1},\ \hbp_{2},\ \hbp_{3}\}$ as the major, intermediate and minor principal directions.
\item
Find at which grid each selected halo is placed. Calculate three cosines of the angles at the grid, $\cth_{i}=\vert\hbs\cdot\hbp_{i}\vert$ 
for $i\in \{1,2,3\}$.
Divide the range of the logarithmic masses of the galactic halos, $m_{h}\equiv \lm$, into multiple differential intervals, 
$[m_{h}, m_{h} +dm_{h}]$ with $dm_{h} = 1$.  Determine the probability densities, $p(\cth_{i})$, by counting the number, $n_{g}$, of the halos  
whose masses fall in each differential interval and take the ensemble averages as $\mcthi \equiv \int_{0}^{1}\,p(\cth_{i})\, \cth_{i}\,d\cth_{i}$ 
at each differential interval.  The error, $\sigma_{\mcthi}$, in the measurement of each ensemble average is calculated as the one standard deviation 
in the mean: $\sigma^{2}_{\mcthi}\equiv [\int_{0}^{1}\,p(\cth_{i})\, (\cth_{i}-\mcthi)^{2}\,d\cth_{i}]/(n_{g}-1)$. 

\end{itemize}

\subsection{Spin Transition Zone}\label{sec:zone}

Figure \ref{fig:eali_all_z0} plots $\langle\cth_{1}\rangle$ (blue line), $\langle\cth_{2}\rangle$ (red line) and $\langle\cth_{3}\rangle$ (green line) 
with one standard deviation errors at $z=0$ as functions of $m_{h}$ for the $\Lambda$CDM (top panel), $w$CDM (middle panel) and RPCDM (bottom panel) models. 
For this plot, the scale radius $R_{f}$ of the Gaussian filter is set at $3\,h^{-1}$Mpc. In each panel, the horizontal black dotted line corresponds to the 
expected value of $\mcthi$ for the case that $\hbs$ is randomly oriented with respect to $\hbp_{i}$. 
For all of the three models, we find that $\mcthb$ ($\mcthc$) is an increasing (decreasing) function of $m_{h}$ while $\mctha$ shows almost no 
variation, being negative constant in the whole range of $11.5\le m_{h}\le 13$, and that the two functions, $\mcthb$ and $\mcthc$ intersect 
each other at a certain threshold mass, signaling the occurrence of the halo spin transition from the $\hbs$-$\hbp_{3}$ alignments 
from the $\hbs$-$\hbp_{2}$ alignments. The three models, however, differ in the rate at which $\mcthb$ ($\mcthc$) increases (decreases) 
with $m_{h}$ as well as in the value of the threshold mass. 

Instead of regarding the threshold mass at which $\mcthb\sim\mcthc$ as the transition mass of the halo spins, however, we adopt the more rigorous 
definition given by \citet{lee-etal20}, according to which the spin transition occurs in a differential mass interval where the null hypothesis of 
$p(\cth_{2})\sim p(\cth_{3})$ is rejected by the Kolmogorov–Smirnov (KS) test at the confidence level lower than $99.9\%$. 
For the KS test of the null hypothesis of $p(\cth_{2})\sim p(\cth_{3})$ at each $m_{h}$-interval, 
we evaluate two cumulative probability functions, $P(\cth_{2}<\cth)\equiv \int_{0}^{\cth}p(\cth_{2})d\cth_{2}$ and 
$P(\cth_{3}<\cth)\equiv \int_{0}^{\cth}p(\cth_{3})d\cth_{3}$. 

\begin{figure}
\begin{center}
\includegraphics[scale=0.7]{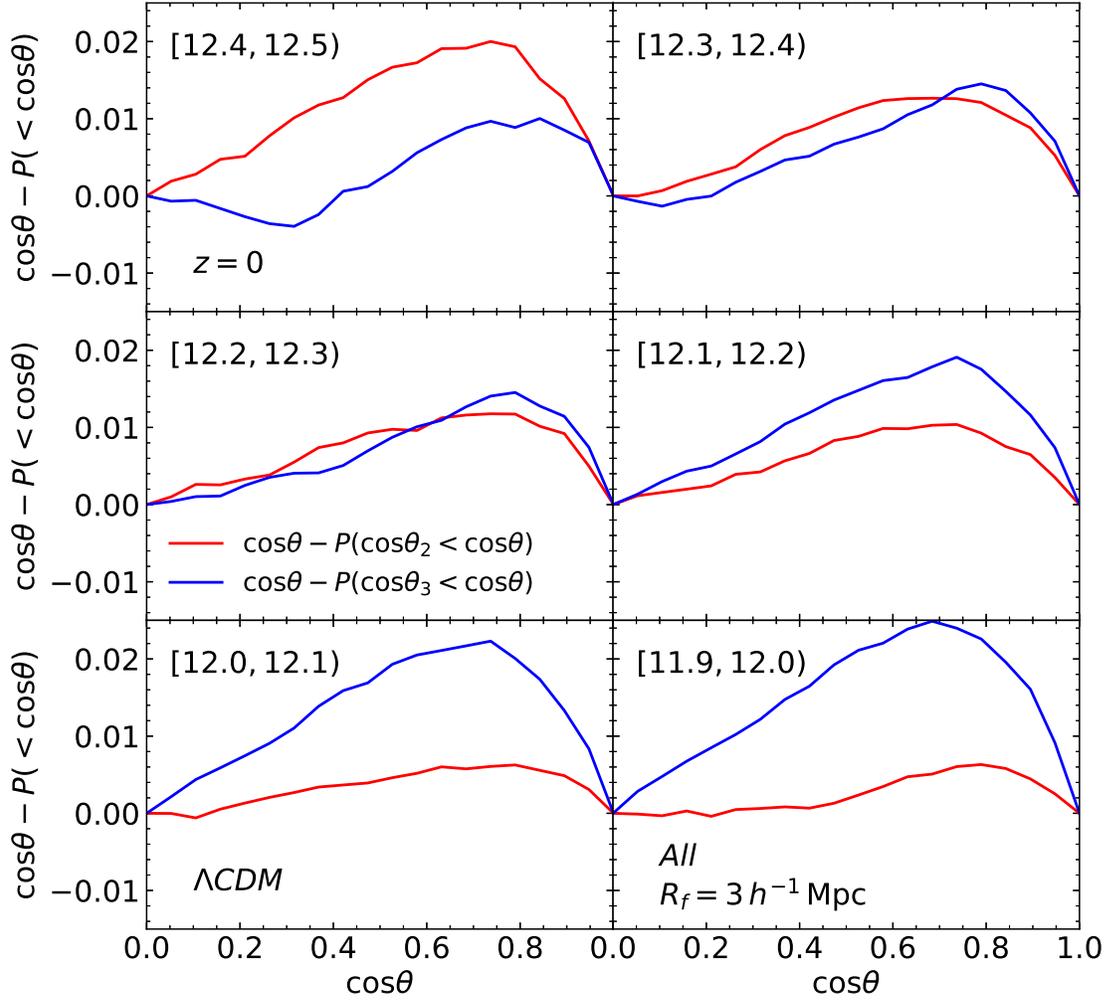}
\caption{Differences between $\cos\theta$ and the cumulative probability distribution 
$P(\cos\theta_{i}<\cos\theta)$ with $i\in \{2,3\}$ at six different mass bins at $z=0$ for the 
$\Lambda$CDM model.}
\label{fig:cbin_all_lcdm_z0}
\end{center}
\end{figure}

\begin{figure}
\begin{center}
\includegraphics[scale=0.7]{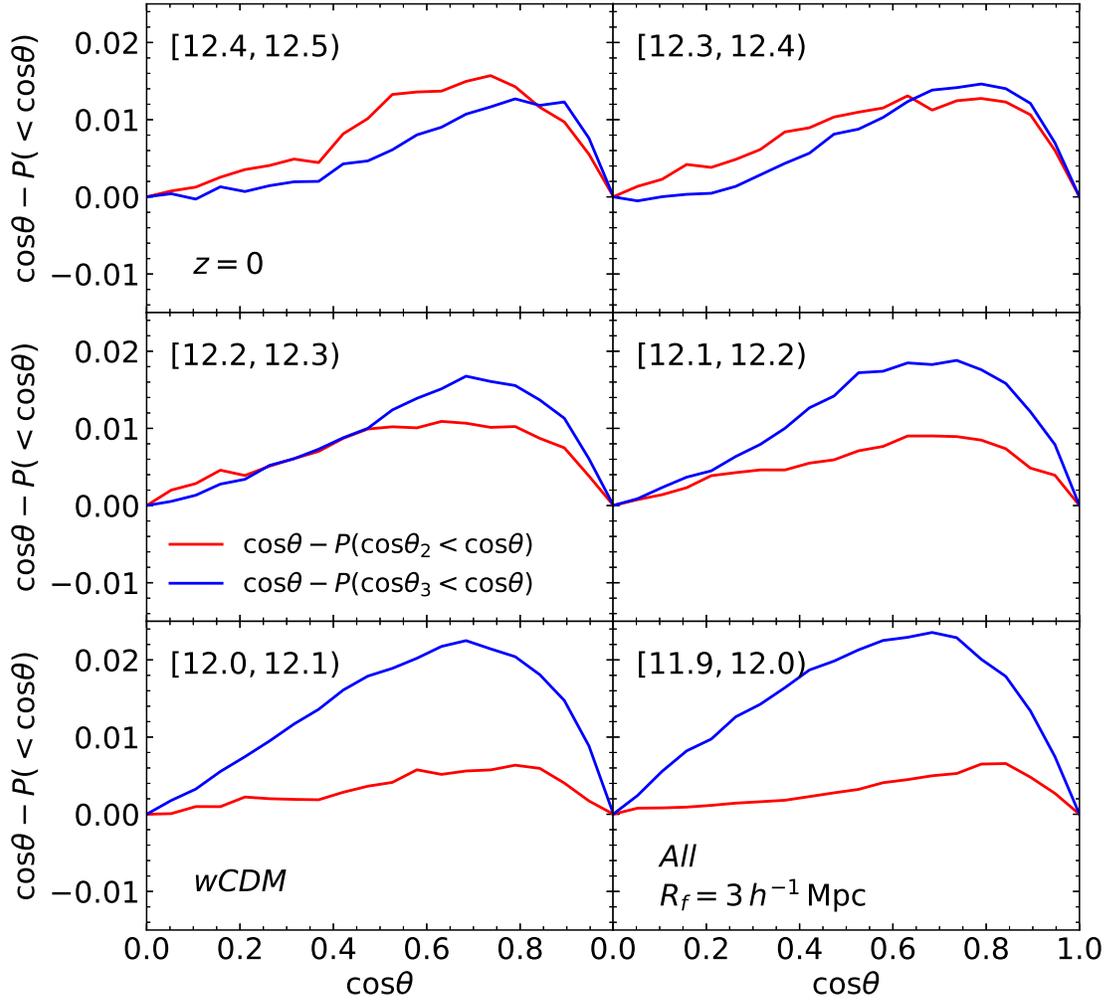}
\caption{Same as Figure \ref{fig:cbin_all_lcdm_z0} but for the $w$CDM model.}
\label{fig:cbin_all_wcdm_z0}
\end{center}
\end{figure}

\begin{figure}
\begin{center}
\includegraphics[scale=0.7]{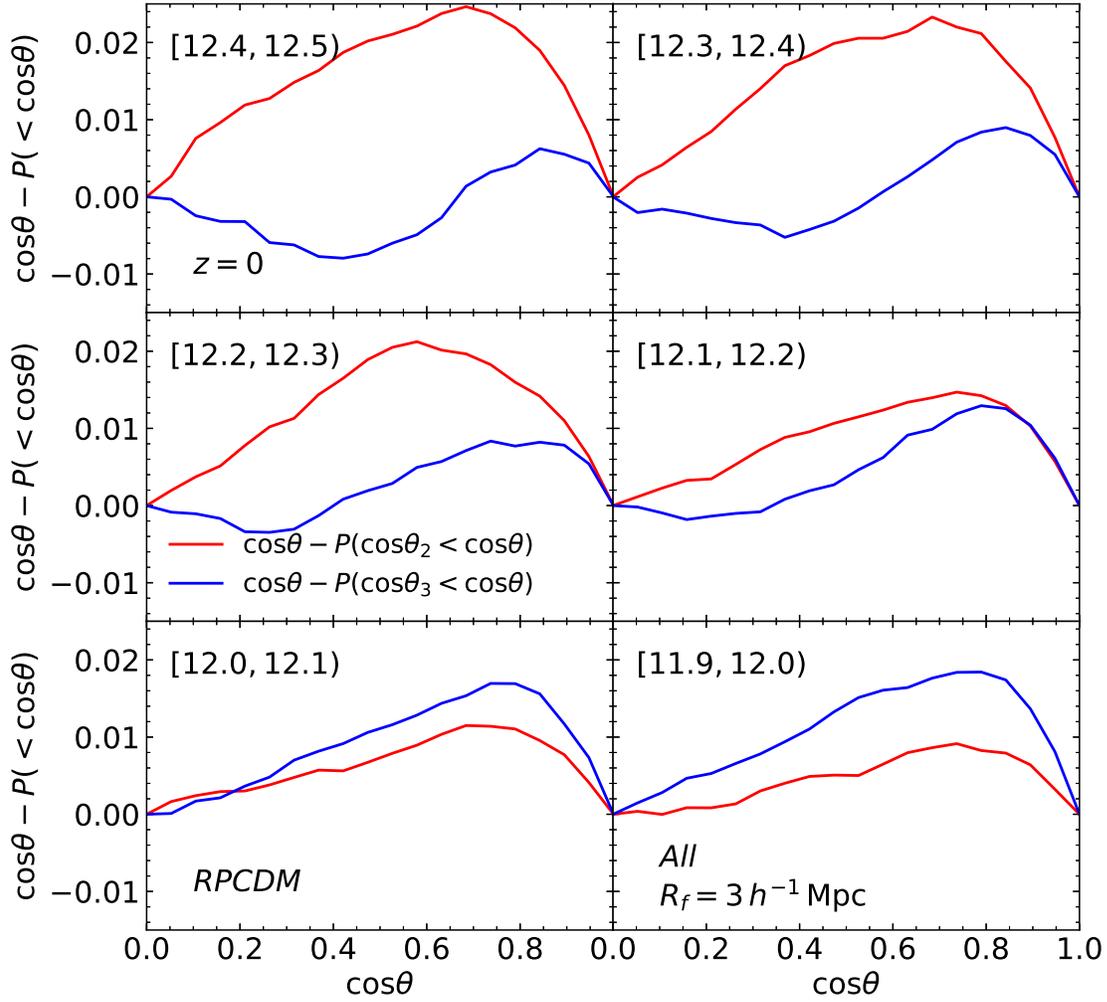}
\caption{Same as Figure \ref{fig:cbin_all_lcdm_z0} but for the RPCDM model.}
\label{fig:cbin_all_rpcdm_z0}
\end{center}
\end{figure}

Figure \ref{fig:cbin_all_lcdm_z0} compares $P(\cth_{2}<\cth)$ (red line) with $P(\cth_{3}<\cth)$ (blue line) at six different 
$m_{h}$-intervals, $12.4\le m_{h}< 12.5$ (top-left panel), $12.3\le m_{h}< 12.4$ (top-right panel),  $12.2\le m_{h}< 12.3$ (middle-left panel),  
$12.1\le m_{h}< 12.2$ (middle-right panel), $12.0\le m_{h}< 12.1$ (bottom-left panel), and $12.0\le m_{h}< 12.1$ (bottom-left panel), 
for the $\Lambda$CDM case at $z=0$.  The same comparisons between the two cumulative distributions but for the $w$CDM and RPCDM cases 
are made in Figures \ref{fig:cbin_all_wcdm_z0} and \ref{fig:cbin_all_rpcdm_z0}, respectively.  To show more clearly the differences between 
$P(\cth_{2}<\cth)$ and $P(\cth_{3}<\cth)$, we plot $\cth-P(\cthi<\cth)$ in lieu of $P(\cthi<\cth)$ in Figures \ref{fig:cbin_all_lcdm_z0}-\ref{fig:cbin_all_rpcdm_z0}, 
as done in \citet{lee-etal20}.

In the differential interval of $12.4\le m_{h}< 12.5$, we find $P(\cth_{2}<\cth)\sim P(\cth_{3}<\cth)$ for the $w$CDM case but  
$P(\cth_{2}<\cth)\nsim P(\cth_{3}<\cth)$ for the other two cases.
In the differential interval of $12.3\le m_{h}< 12.4$, we find $P(\cth_{2}<\cth)\sim P(\cth_{3}<\cth)$ for the $w$CDM and $\Lambda$CDM cases, 
but $P(\cth_{2}<\cth)\nsim P(\cth_{3}<\cth)$ for the RPCDM case.
In the differential interval of $12.0\le m_{h}< 12.1$, we find $P(\cth_{2}<\cth)\sim P(\cth_{3}<\cth)$ for the RPCDM case, 
but $P(\cth_{2}<\cth)\nsim P(\cth_{3}<\cth)$ for the $w$CDM and $\Lambda$CDM cases.
These results clearly indicate that the three DE models differ in the mass intervals where $p(\cth_{2})\sim p(\cth_{3})$.
 
We compute the maximum distance, $D_{\rm max}$, between $P(\cth_{2}<\cth)$ and $P(\cth_{3}<\cth)$ at each $m_{h}$-interval and multiply it by the weighting factor 
$\sqrt{n_{h}/2}$, where $n_{h}$ denotes the number of the halos in each $m_{h}$-interval.  Note that both of $D_{\rm max}$ and $n_{h}$ are functions of $m_{h}$, i.e., 
$D_{\rm max}=D_{\rm max}(m_{h})$ and $n_{h}=n_{h}(m_{h})$.  
If this weighted maximum distance turns out to be lower (higher) than the  critical value, $1.949$, at a given $m_{h}$-interval, then the KS test rejects the 
null hypothesis of $p(\cth_{2})\sim p(\cth_{3})$ at the confidence level lower (higher) than $99.9\%$.  
Introducing a new concept, the {\it spin transition zone}, $\Delta m_{t}$, defined as
\begin{equation}
\label{eqn:kstest}
\Delta m_{t}=\bigg{\{}m_{h}\vert \sqrt{\frac{1}{2}n_{h}(m_{h})}D_{\rm max}(m_{h})<1.949\bigg{\}}\, ,
\end{equation}
we take into account the fact that the halo spin transition in fact does not occur sharp at a singular threshold mass scale but gradually proceed in a finite 
mass range.

Figure \ref{fig:cl_all_z0} plots the weighted maximum distance, $\sqrt{n_{h}/2}D_{\rm max}$, between $P(\cth_{2}<\cth)$ and $P(\cth_{3}<\cth)$ 
as a function of $m_{h}$ for the three models. 
In each panel, the horizontal black dashed line corresponds to the critical value, $1.949$ for the $99.9\%$ confidence level according to the KS test. 
As can be seen, the three DE models have significantly different spin transition zones: $\mt=\{m_{h}\vert 12.2\le m_{h}< 12.4\}$ ($\Lambda$CDM), 
$\mt=\{m_{h}\vert 12.2\le m_{h}< 12.5\}$ ($w$CDM), and $\mt=\{m_{h}\vert 12.0\le m_{h}< 12.1\}$ (RPCDM). The spin transition zone of the RPCDM 
model turns out to be most conspicuously different from those of the other two models, being biased toward the significantly lower mass section. 
Although the spin transition zone of the $\Lambda$CDM model is overlapped with that of the $w$CDM model, the two models can still be distinguished by 
the non-overlapped $m_{h}$-interval, $12.4\le m_{h}\le 12.5$,  where the null hypothesis is rejected by the KS test at the confidence level higher than $99.9\%$ 
for the $\Lambda$CDM case but not for the $w$CDM case. In other words, the halos with masses in the range of $12.4\le m_{h}\le 12.5$ do not show a spin 
transition for the $\Lambda$CDM case, while they do for the $w$CDM case. 

Smoothing the tidal shear field on a larger scale $R_{f}=5\dunit$, we recalculate $\mcthi$ and $\sqrt{n_{h}/2}D_{\rm max}$, which 
are depicted in Figures \ref{fig:eali_all_rf5_z0} and \ref{fig:cl_all_rf5_z0}, respectively. As can be seen, the variation of $R_{f}$ to $5\dunit$ 
does not weaken the DE dependence of $\mt$, although it has an effect of slightly widening $\mt$ for all of the three models. 
We find $\mt=\{m_{h}\vert 12.3\le m_{h}< 12.6\}$ ($\Lambda$CDM), $\mt=\{m_{h}\vert 12.3\le m_{h}< 12.5\}$ ($w$CDM), and 
$\mt=\{m_{h}\vert 12.1\le m_{h}< 12.3\}$ (RPCDM). The RPCDM model can still be readily distinguished from the other two models by 
its spin transition zone in the lowest mass section.  
The $\Lambda$CDM and $w$CDM models still significantly differ from each other in the non-overlapped interval of $12.5\le m_{h}< 12.6$ 
where the null hypothesis of $p(\cth_{2})\sim p(\cth_{3})$ is rejected at the confidence level higher (lower) than $99.9\%$ levels 
for the latter (former) model.  

Using the halo catalog at $z=0.4$ from the DEUS and resetting $R_{f}$ at $3\dunit$, we go through the same routine to determine 
$\mcthi$ and $\sqrt{n_{h}/2}D_{\rm max}$ at $z=0.4$, which are depicted in Figures \ref{fig:eali_all_z0.4} and \ref{fig:cl_all_z0.4}, respectively. 
As can be seen, at higher redshifts, the spin transition zones become narrower, showing considerable shifts toward the lower mass sections 
for all of the three models. The spin transition zones at $z=0.4$ are found to be $\mt=\{m_{h}\vert 12.0\le m_{h}< 12.1\}$ ($\Lambda$CDM), 
$\mt=\{m_{h}\vert 12.0\le m_{h}< 12.2\}$ ($w$CDM), and $\mt=\{m_{h}\vert 11.8\le m_{h}< 11.9\}$ (RPCDM). Note that the spin transition 
zone of the RPCDM model is located in the mass section below $10^{12}\munit$. In other words, the halos with masses equal to or higher than 
$10^{12}\dunit$ do not show a spin transition at $z=0.4$ in the RPCDM model.  The $\Lambda$CDM model can still be distinguished from the 
$w$CDM model at $z=0.4$ by  the non-overlapped $m_{h}$-interval of $12.1\le m_{h}< 12.2$ where the confidence level for the rejection of the 
null hypothesis drops below (stays above) $99.9\%$ for the $w$CDM ($\Lambda$CDM) case. 

\subsection{Dependence on the Web-Type}\label{sec:web}

\begin{figure}
\begin{center}
\includegraphics[scale=0.7]{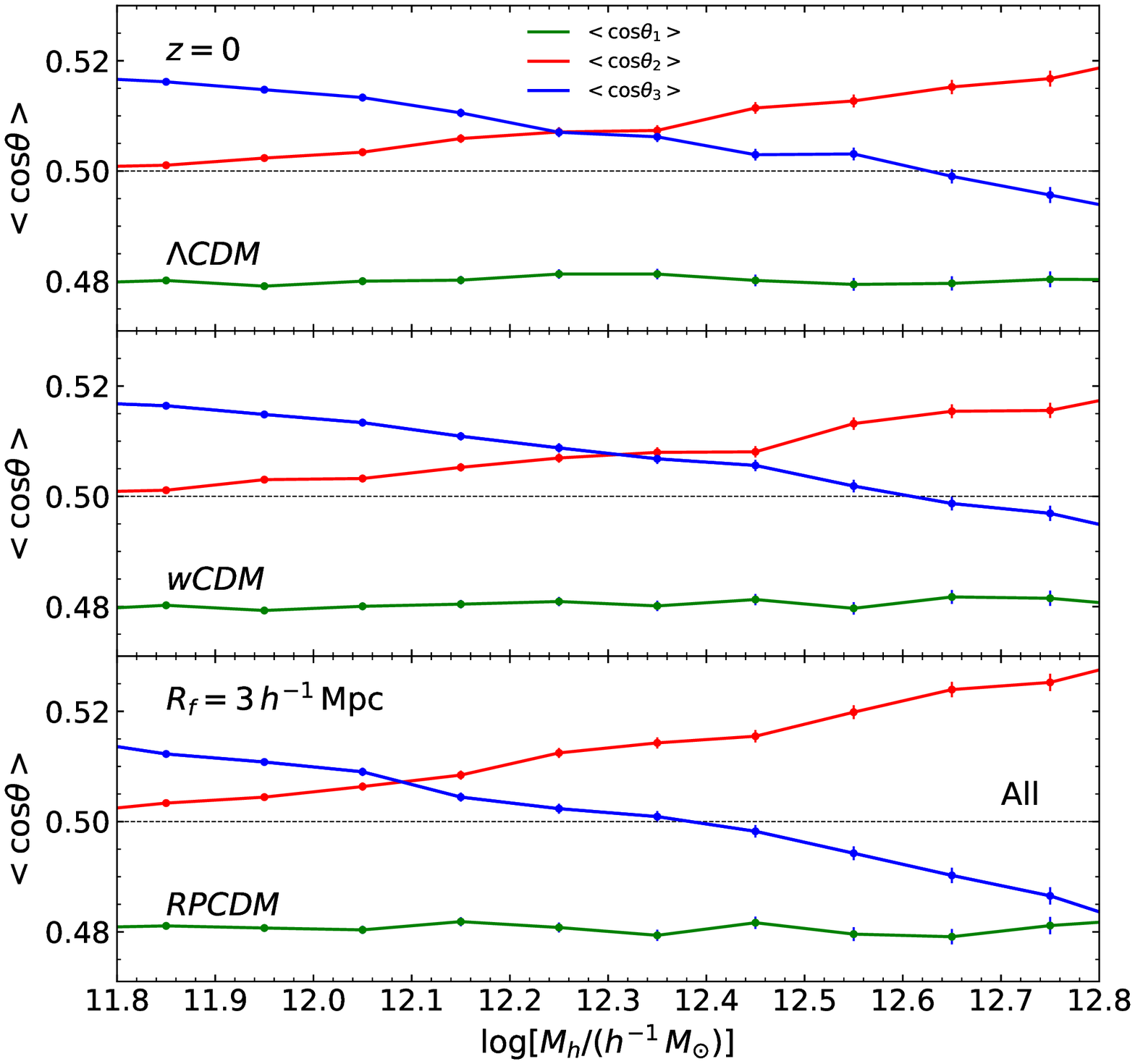}
\caption{Ensemble average of the cosines of the angles between the unit spin vectors of DM halos and 
three principal directions of the local tidal fields at $z=0$ as a function of the halo mass for three different dark 
energy models.}
\label{fig:eali_all_z0}
\end{center}
\end{figure}

\begin{figure}
\begin{center}
\includegraphics[scale=0.7]{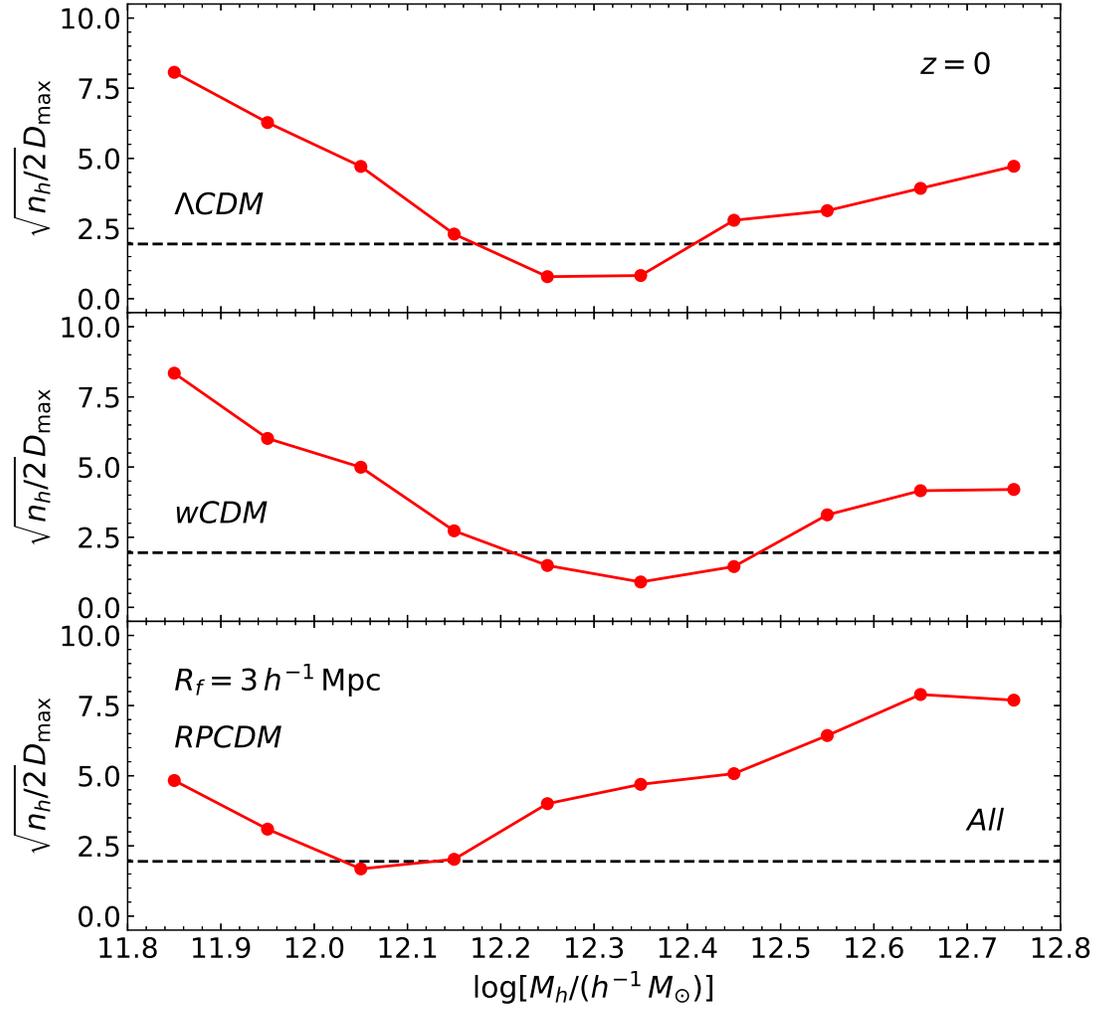}
\caption{Maximum distances between $P(\cth_{2}<\cth)$ and $P(\cth_{3}<\cth)$ multiplied by $\sqrt{n_{h}/2}$  
as a function of the halo mass at $z=0$ for the three DE models.}
\label{fig:cl_all_z0}
\end{center}
\end{figure}

\begin{figure}
\begin{center}
\includegraphics[scale=0.7]{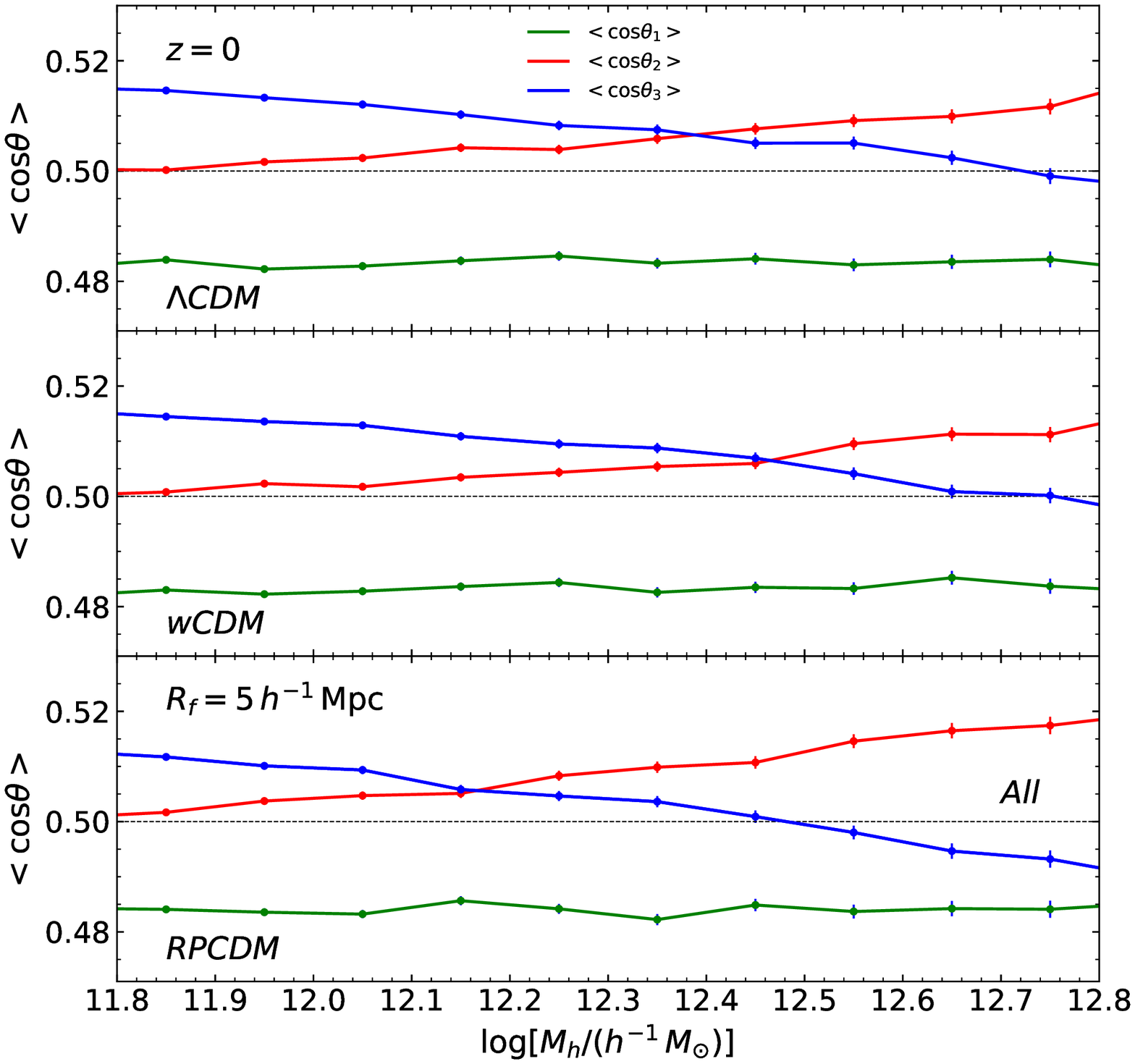}
\caption{Same as Figure \ref{fig:eali_all_z0} but for the case of $R_{f}=5\dunit$.}
\label{fig:eali_all_rf5_z0}
\end{center}
\end{figure}

\begin{figure}
\begin{center}
\includegraphics[scale=0.7]{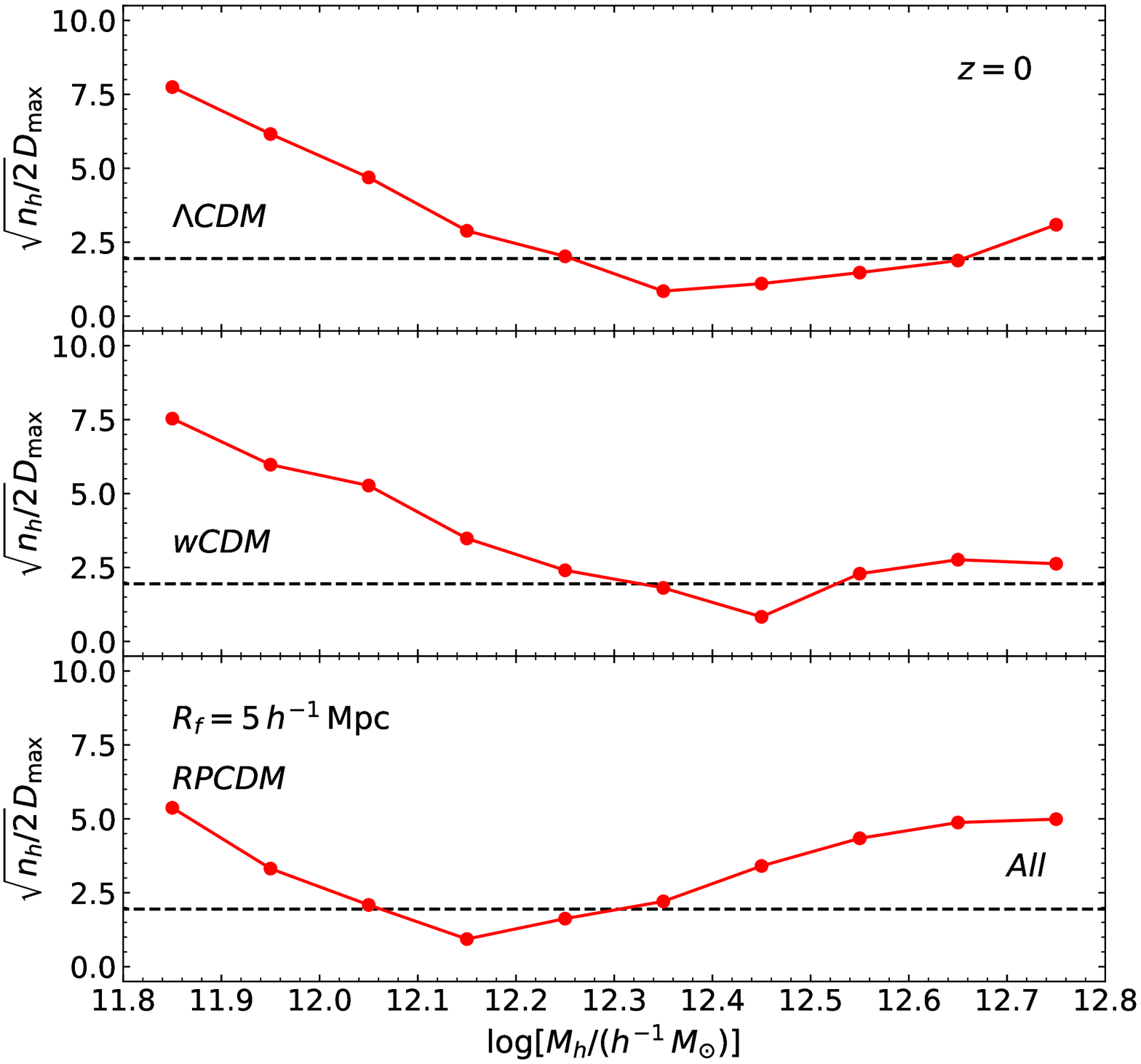}
\caption{Same as Figure \ref{fig:cl_all_z0} but for the case of $R_{f}=5\dunit$.}
\label{fig:cl_all_rf5_z0}
\end{center}
\end{figure}

\begin{figure}
\begin{center}
\includegraphics[scale=0.7]{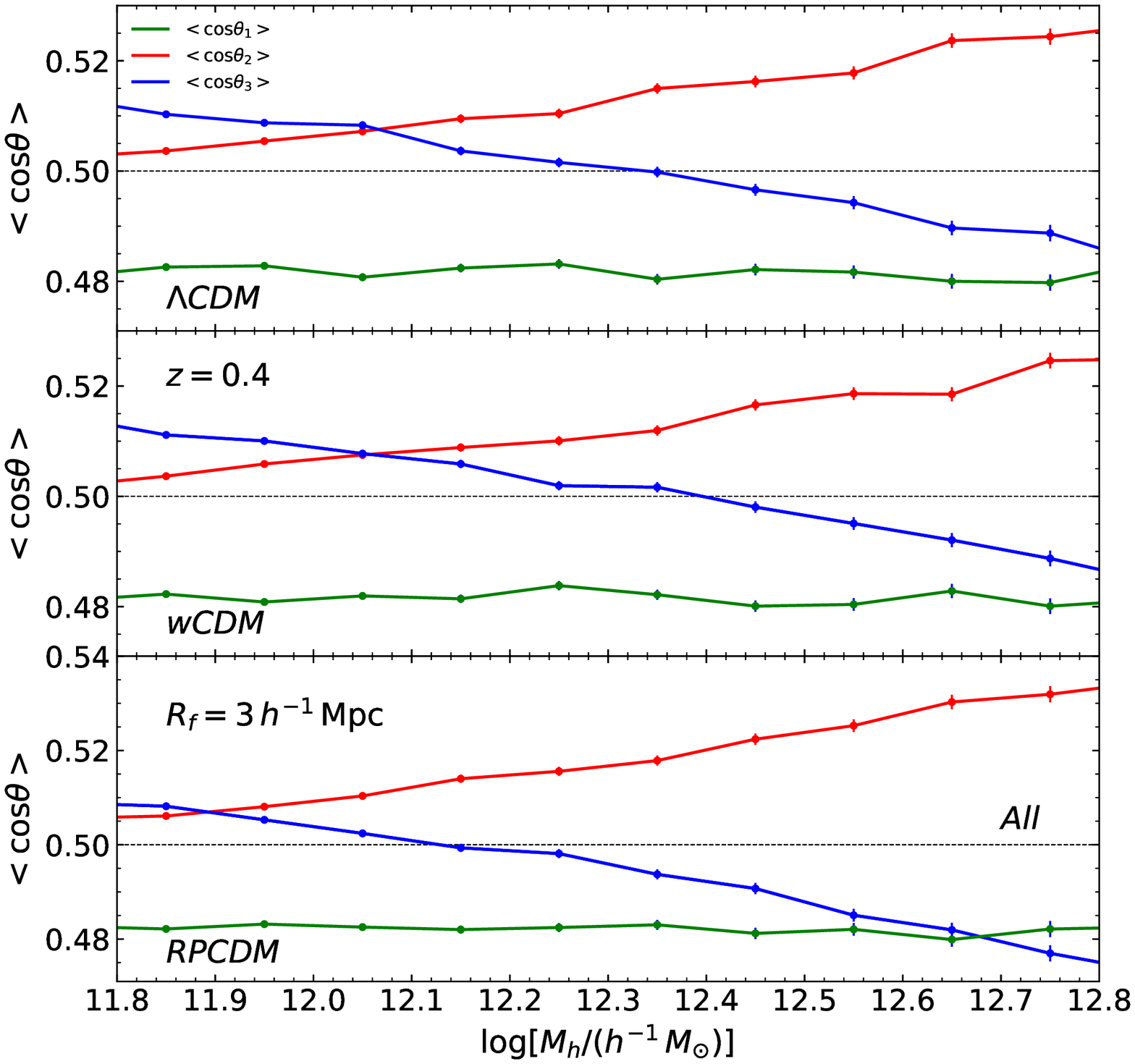}
\caption{Same as Figure \ref{fig:eali_all_z0} but at $z=0.4$.}
\label{fig:eali_all_z0.4}
\end{center}
\end{figure}

\begin{figure}
\begin{center}
\includegraphics[scale=0.7]{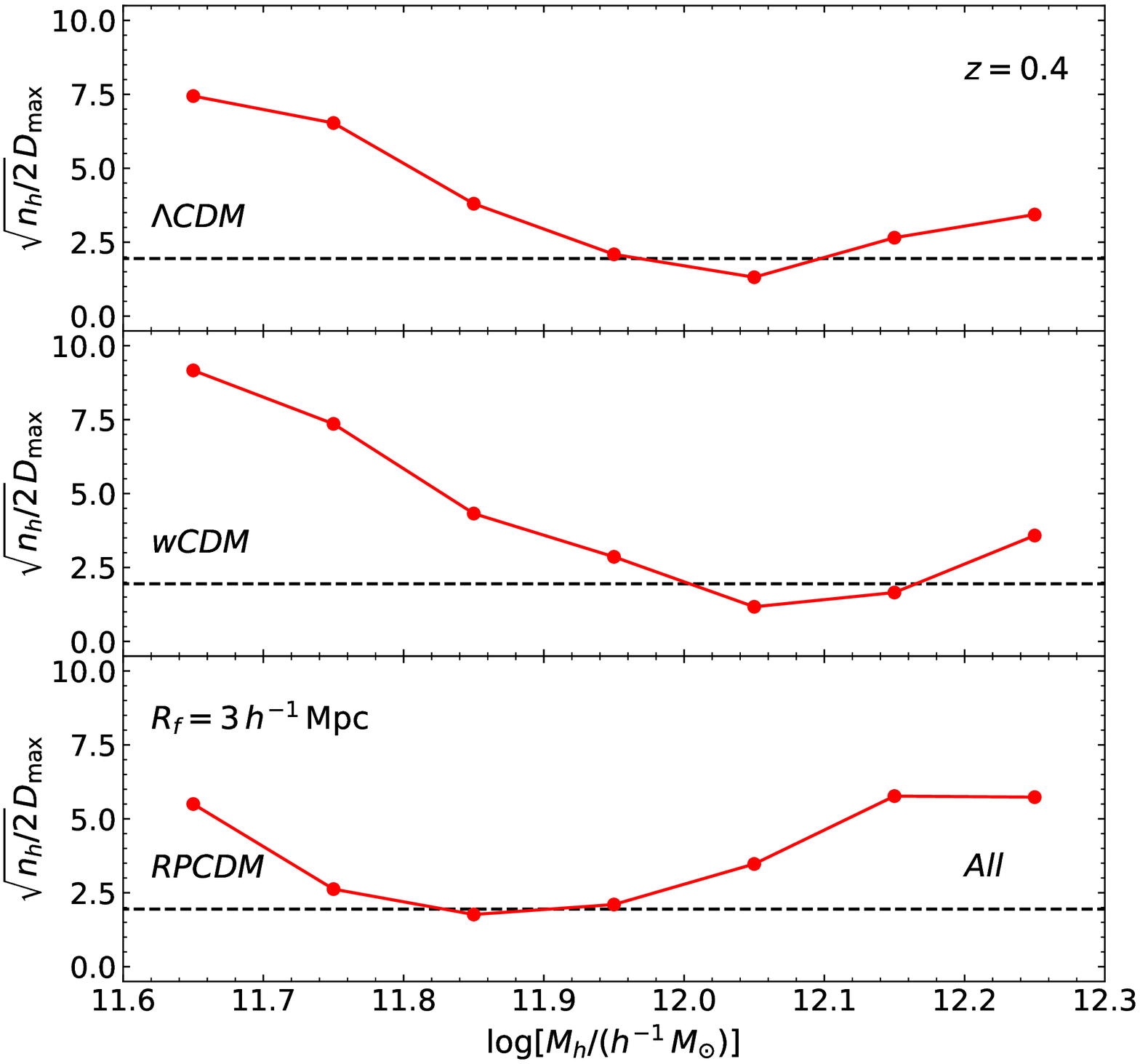}
\caption{Same as Figure \ref{fig:cl_all_z0} but at $z=0.4$.}
\label{fig:cl_all_z0.4}
\end{center}
\end{figure}

\begin{figure}
\begin{center}
\includegraphics[scale=0.7]{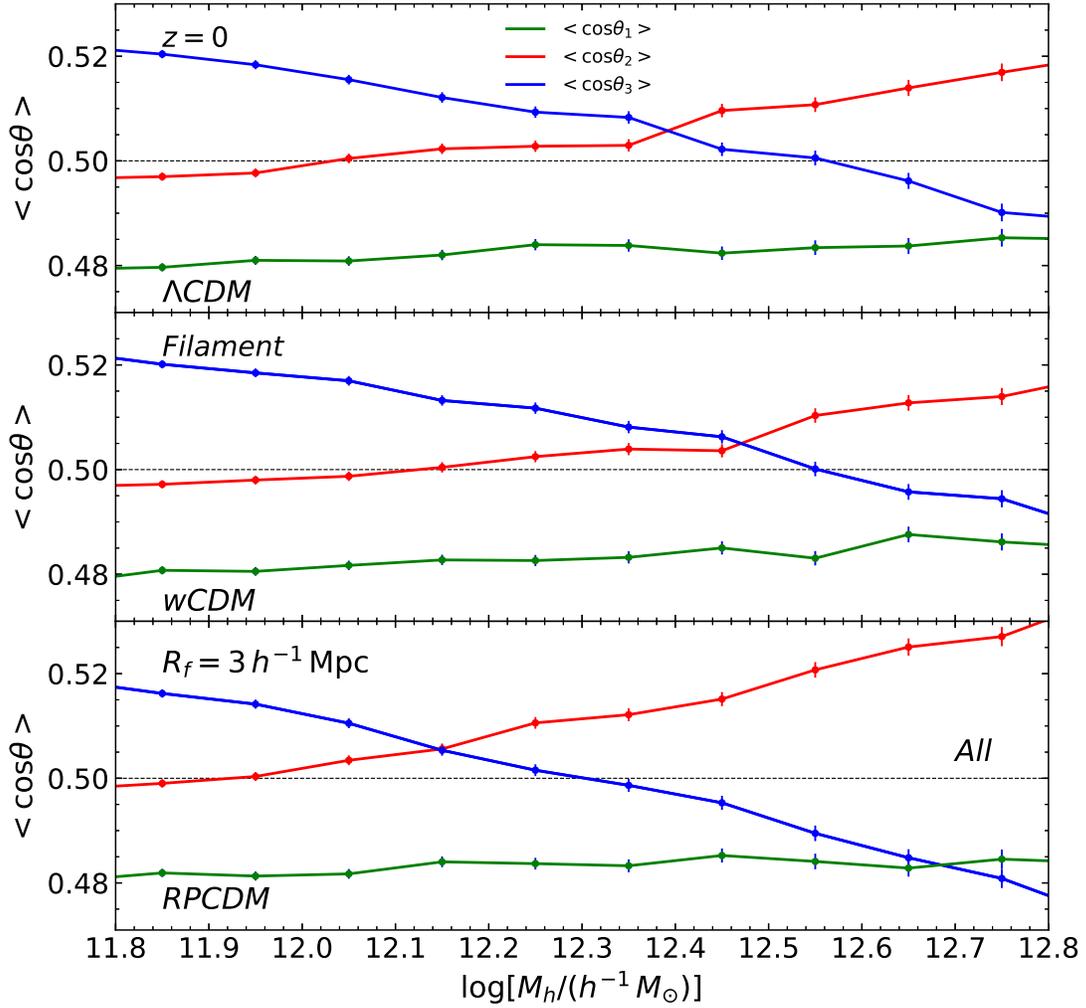}
\caption{Same as Figure \ref{fig:eali_all_z0} but only with the filament halos.}
\label{fig:eali_fil_z0}
\end{center}
\end{figure}

\begin{figure}
\begin{center}
\includegraphics[scale=0.7]{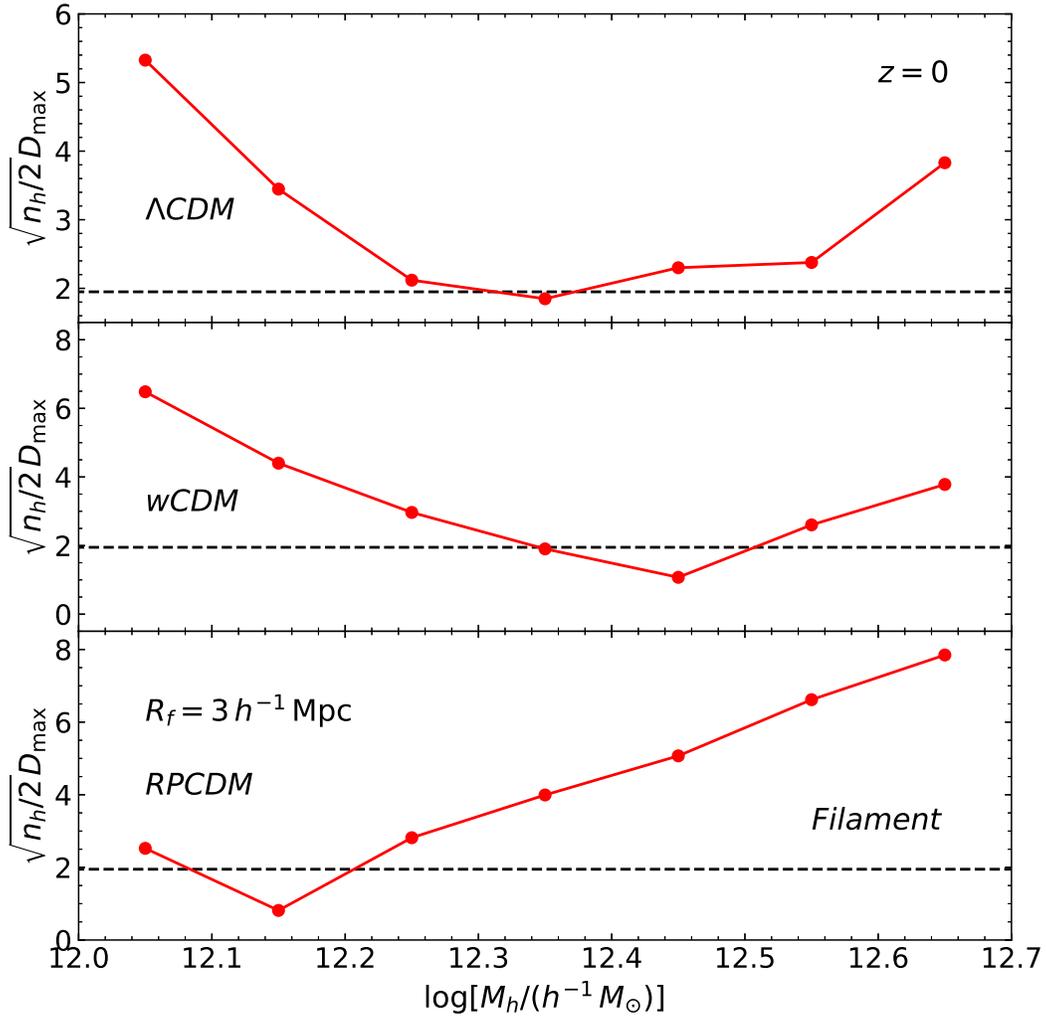}
\caption{Same as Figure \ref{fig:cl_all_z0} but for the case of the filament halos.}
\label{fig:cl_fil_z0}
\end{center}
\end{figure}

To investigate how the spin transition zone of each model varies with the web type, we classify the halo environments into the four web 
types, namely, knots, filaments, sheets and voids according to the signs of the tidal eigenvalues, $\{\lam_{i}\}_{i=1}^{3}$ at the grids where 
each galactic halo is placed \citep{hah-etal07a}. 
Figure \ref{fig:eali_fil_z0} (Figure \ref{fig:cl_fil_z0}) plots the same as Figure \ref{fig:eali_all_z0} (Figure \ref{fig:cl_all_z0}) but using only those halos 
embedded in the filaments satisfying the condition of $\lam_{1}\ge\lam_{2}>0$ and $\lam_{3}<0$. The spin transition zones of the filament 
halos are determined to be $\mt=\{m_{h}\vert 12.3\le m_{h}< 12.4\}$ ($\Lambda$CDM), 
$\mt=\{m_{h}\vert 12.4\le m_{h}< 12.5\}$ ($w$CDM), and $\mt=\{m_{h}\vert 12.1\le m_{h}< 12.2\}$ (RPCDM).
Comparing Figure \ref{fig:cl_fil_z0} with Figure \ref{fig:cl_all_z0}, we note that the spin transitions of the filament halos occur in relatively narrow 
mass intervals for all of the three models and that in the filaments the spin transition zones between the $\Lambda$CDM and the $w$CDM cases 
are no longer overlapped, which explains why the filaments are the most optimal environments for the study of the halo spin transitions.

\begin{figure}
\begin{center}
\includegraphics[scale=0.7]{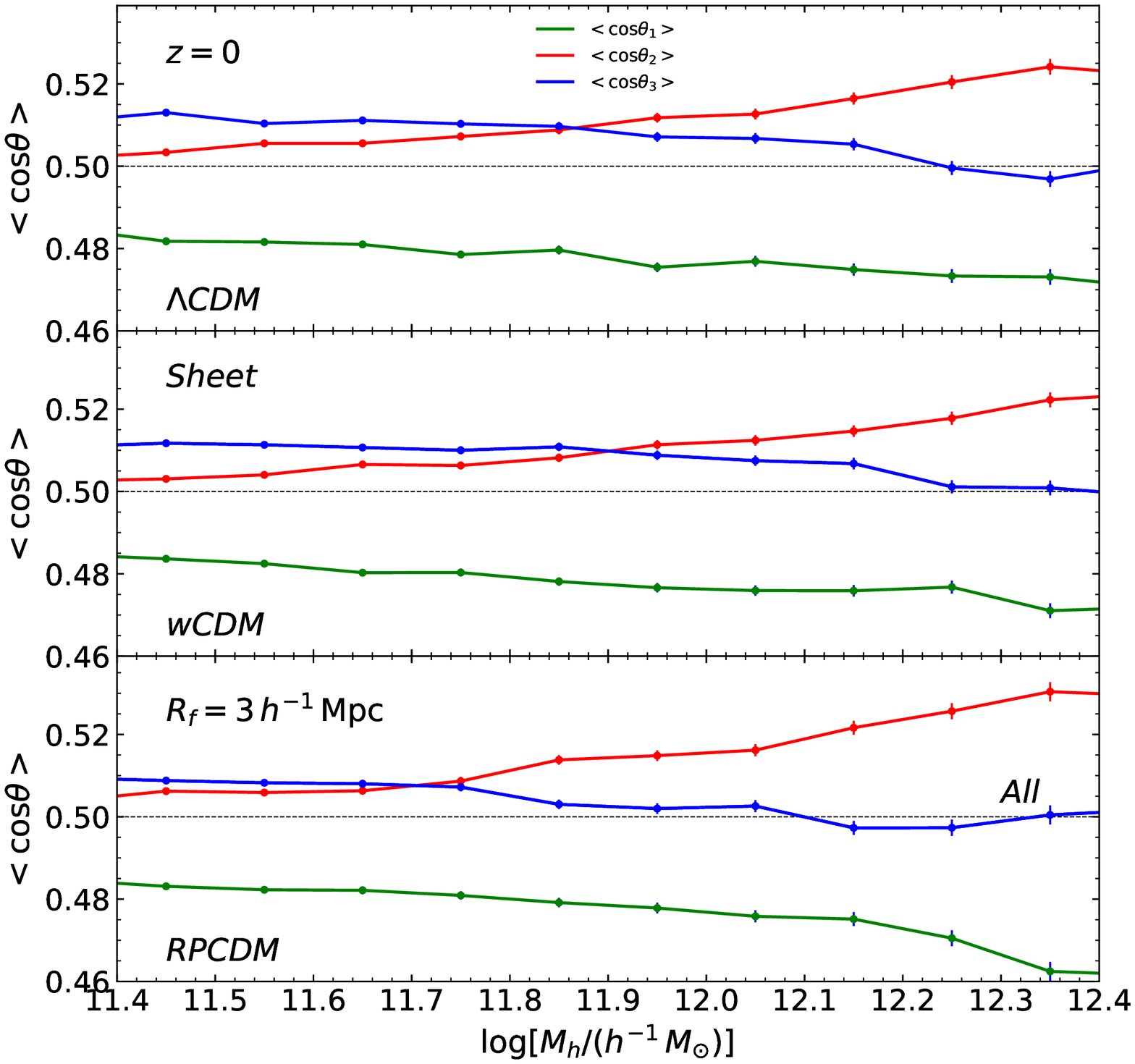}
\caption{Same as Figure \ref{fig:eali_all_z0} but only with the sheet halos.}
\label{fig:eali_sheet_z0}
\end{center}
\end{figure}

\begin{figure}
\begin{center}
\includegraphics[scale=0.7]{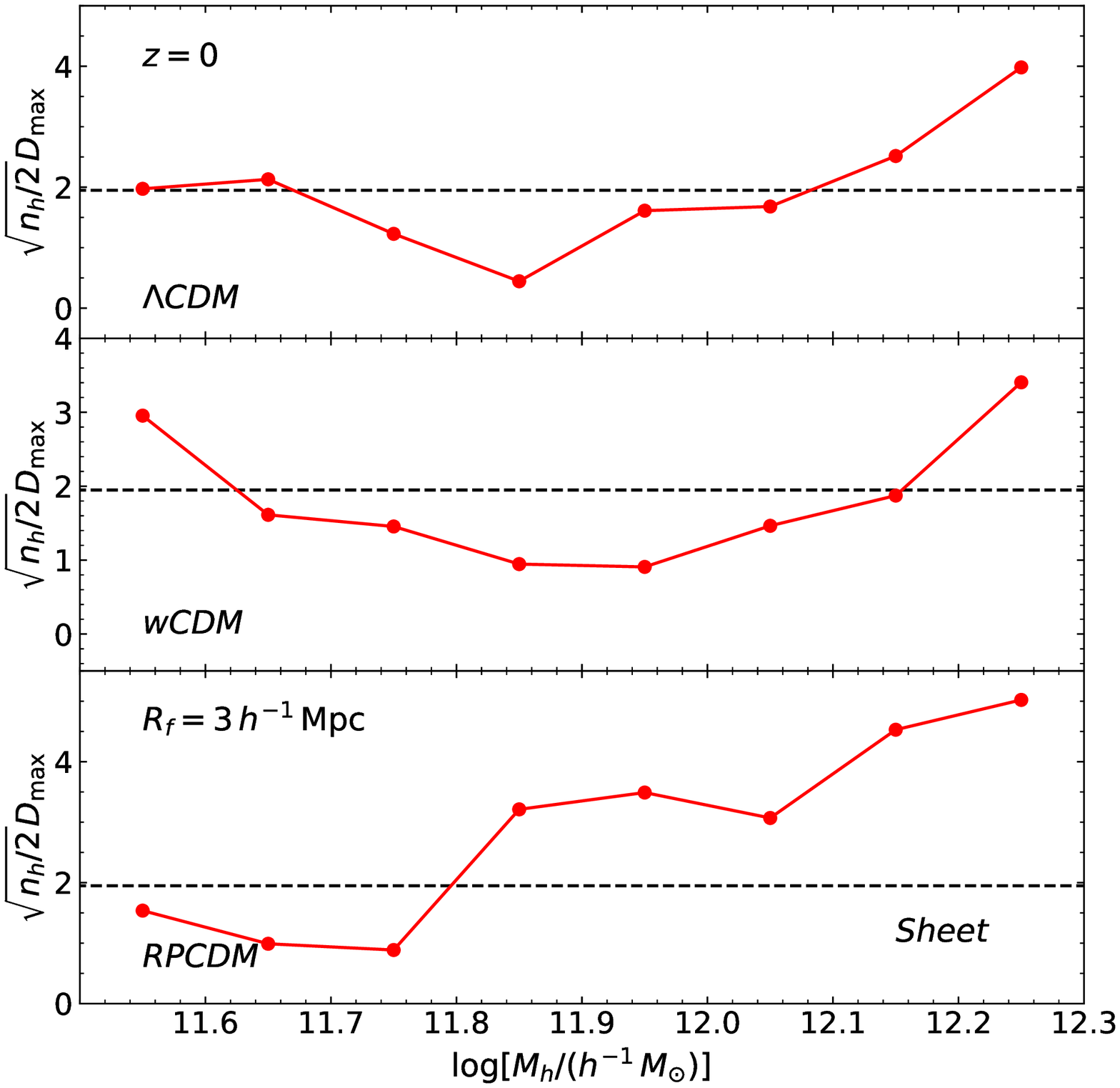}
\caption{Same as Figure \ref{fig:cl_all_z0} but only with the sheet halos.}
\label{fig:cl_sheet_z0}
\end{center}
\end{figure}

Figure \ref{fig:eali_sheet_z0} (Figure \ref{fig:cl_sheet_z0}) plots the same as Figures \ref{fig:eali_all_z0} (Figure \ref{fig:cl_all_z0}) but using only those 
halos in the sheets satisfying the condition of $\lam_{1}>0$ and $\lam_{2}<0$.  As can be seen, the sheet halos exhibit the spin transitions, too 
\citep{WK17}. Their spin transition zones are, however, shifted to the lower mass sections compared with those of the filament halos for all of the three models: 
$\mt=\{m_{h}\vert 11.7\le m_{h}< 12.1\}$ ($\Lambda$CDM), $\mt=\{m_{h}\vert 11.6\le m_{h}< 12.2\}$ ($w$CDM), and 
$\mt=\{m_{h}\vert 11.5\le m_{h}< 11.8\}$ (RPCDM).  
Comparing Figure \ref{fig:cl_sheet_z0} with Figure \ref{fig:cl_all_z0}, we also note that the spin transitions of the sheet halos occur in relatively broad 
mass intervals for all of the three models and that in the sheets the spin transition zones are overlapped not only between the $\Lambda$CDM and 
$w$CDM cases but also between the RPCDM and $w$CDM cases, which indicates that the sheets are not so optimal as the filaments for the 
investigation of $\mt$ as a discriminator of DE models. 
It is, however, worth mentioning here that for the $w$CDM and RPCDM cases the sheet halos exhibit the spin transitions in the mass range 
lower than $\log [M_{h}/(\munit)]<11.8$ where the directions of the halo spin vectors are likely to be inaccurate due to the low particle resolution 
\citep{bet-etal07}.  For a more accurate investigation of the spin transition of the sheet halos,  it will be necessary to use data from a higher-resolution 
N-body simulation.

We have also determined the spin transition zones of the halos embedded in the knots ($\lam_{3}>0$) and voids ($\lam_{1}<0$) for each model 
and found that the three DE models cannot be discriminated from one another by the spin transition zones of the knot halos nor by the void halos 
due to the large uncertainties. We leave out the results for the cases of the knot and void halos.  

\section{Discussion and Conclusion}\label{sec:con}

In light of the recent finding of \citet{lee-etal20} that the halo spin transitions can be a powerful probe of the total neutrino mass, we have investigated 
whether or not it can be also used to discriminate non-standard DE models from the standard $\Lambda$CDM cosmology. 
For this investigation, we appropriated the halo catalogs from the DEUS which were performed for different DE models, $\Lambda$CDM, $w$CDM 
and RPCDM, which satisfy equally well the constraints from the WMAP7 and SN Ia observations (see Table \ref{tab:initial}) \citep{deus1}.  
The probability density functions of the cosines of the angles between the halo spin axes and the intermediate (minor) principal axes of the local tidal fields, 
$p(\cth_{2})$ ($p(\cth_{3})$), have been numerically evaluated for each DE model.
Adopting the new definition introduced by \citet{lee-etal20},  we have determined for each model a spin transition zone, $\mt$, as the halo mass range in 
which the confidence level for the rejection of the null hypothesis of $p(\cth_{2})\sim p(\cth_{3})$ by the KS test descends below $99.9\%$. 

We have shown that the three DE models significantly differ in $\mt$ among themselves. The summary of our results is the following.
\begin{itemize}
\item
At $z=0$, the spin transition zone of the RPCDM model resides in the lowest mass section among the three, which allows itself to be plainly distinguished 
from the $\Lambda$CDM model. Although the spin transition zone of the $w$CDM model is partially overlapped with that of $\Lambda$CDM,  the 
$w$CDM model can still be discriminated with high statistical significance from the $\Lambda$CDM model by the largest halo mass contained in the spin 
transition zone at $z=0$.
\item
At higher redshifts $z=0.4$, the overall strengths of the halo spin alignments with the tidal principal directions for all of the three models are enhanced, 
which leads the spin transition zone to be shifted toward the lower mass section, enlarging the differences in $\mt$ among the three models. 
\item
The increase of the smoothing scale of the tidal fields undermines the alignment tendency of the halo spins with the tidal intermediate principal 
directions, shifting the spin transition zone toward the higher mass section for all of the three models. 
Nevertheless, the statistical significance of the differences in $\mt$ among the three dark energy models is robust against the 
variations of the smoothing scales.  
\item
In the filaments, the strengths of the halo spin alignments with the tidal principal axes vary relatively rapidly with the halo mass, which leads the spin 
transition to occur in a narrower mass range (i.e., narrower spin transition zone) for all of the three models. In consequence, the differences in the spin 
transition zone among the three models become more prominent in the filaments. 
\item
In the sheets, the strengths of the halo spin alignments with the tidal principal axes vary relatively slowly with the halo mass, which has an effect of 
widening the spin transition zone, shifting it to the lower mass section \citep[cf.][]{WK17}. In consequence, the differences in the spin transition zone among 
the three models become less prominent in the sheets. 
\end{itemize}

Our results is in line with the claim of \citet{lee-etal20} that the spin transition mass depends on how fast the nonlinear evolution of the tidal fields proceed. 
According to the linear tidal torque theory \citep{whi84}, the spin axes of the galactic halos, $\hbs$, prefer the intermediate principal directions, $\hbp_{2}$, 
of the linear tidal fields before the turn-around moments \citep{LP00,LP01}.  The subsequent nonlinear evolution of the tidal fields, however, gradually 
deviates their principal axes from the original directions, which lessens the linearly generated $\hbs$-$\hbp_{2}$ alignments and simultaneously promotes the 
$\hbs$-$\hbp_{3}$ alignments. 
According to this claim, the spin transition zone would shift toward the higher (lower) mass section in a model where the tidal field undergoes a faster 
(slower) nonlinear evolution.  
Given that the ratio of the linear growth factor in the RPCDM ($w$CDM) model to that of the $\Lambda$CDM model is always below (above) unity 
\citep{deus1},  the nonlinear evolution of the tidal field in the RPCDM ($w$CDM) model is expected to proceed more slowly (rapidly) than that in the $\Lambda$CDM 
model, which explains why the spin transition in the former model occurs in the lower (higher) mass section than that of the latter model. 

The bottom line is that the spin transition zone can in principle discriminate the non-standard DE models from the standard $\Lambda$CDM model. 
The advantage of using the spin transition zone as a DE discriminator over the linear growth factor is that the spin transition zone can be estimated at 
the present epoch from real observational data, while the high-$z$ data are required to estimate the linear growth factor. Moreover, recalling that 
the spin transition zone deals with more readily observable small-scale features of the universe, we speculate that it might contain additional information 
on the nonlinear evolution of structure formation, having a potential to break the other cosmic degeneracies.  
Our future work is in the direction of extending the current analysis to a broader range of non-standard cosmologies and to more comprehensively study 
the efficacy of the halo spin transition zone as a probe of cosmology.

\acknowledgments

J.L. acknowledge the support by Basic Science Research Program through the National Research Foundation (NRF) of Korea 
funded by the Ministry of Education (No.2019R1A2C1083855) and also by a research grant from the NRF to the Center for Galaxy 
Evolution Research (No.2017R1A5A1070354). NIL acknowledges financial support of the Project IDEXLYON at the University of Lyon 
under the Investments for the Future Program (ANR-16-IDEX-0005). 
NIL also acknowledges support from the joint Sino-German DFG research Project ``The Cosmic Web and its impact on galaxy formation 
and alignment'' (DFG-LI 2015/5-1).

\clearpage
\clearpage
\clearpage
\clearpage
\clearpage
\clearpage
\clearpage
\clearpage
\clearpage
\clearpage
\clearpage
\clearpage
\begin{deluxetable}{ccccccc}
\tablewidth{0pt}
\setlength{\tabcolsep}{3mm}
\tablecaption{Initial conditions and the particle mass resolution}
\tablehead{model & $\Omega_{m}$ & h & $\sigma_{8}$ & $w_{0}$ & $w_{a}$ & $m_{p}$\\  
& & & & & & [$10^{9}\munit$] }
\startdata
$\Lambda$CDM & 0.257 & 0.72 & 0.80 & -1.0 & 0.0 & 2.3 \\
$w$CDM & 0.275 & 0.72 & 0.852 & -1.2 & 0.0  & 2.4 \\
RPCDM & 0.230 & 0.72 & -0.66 & -0.87 & 0.08 & 2.0
\enddata
\label{tab:initial}
\end{deluxetable}
\clearpage
\begin{deluxetable}{ccccc}
\tablewidth{0pt}
\setlength{\tabcolsep}{3mm}
\tablecaption{Halo Spin Transition Zones}
\tablehead{model & web type & $R_{f}$ & $z$ & $\mt$ 
\\  & & [$\dunit]$ &  &}
\startdata
$\Lambda$CDM & all & $3$ & 0 & $[12.2,\ 12.4]$   \\
$w$CDM & all & $3$ &  0 & $[12.2,\ 12.5]$  \\
RPCDM & all & $3$ &  0 & $[12.0,\ 12.1]$  \\
\hline
$\Lambda$CDM & all & $5$ & 0 & $[12.3,\ 12.6]$   \\
$w$CDM & all & $5$ &  0 & $[12.3,\ 12.5]$  \\
RPCDM & all & $5$ &  0 & $[12.1,\ 12.3]$  \\
\hline
$\Lambda$CDM & all & $3$ & 0.4 & $[12.0,\ 12.1]$   \\
$w$CDM & all & $3$ &  0.4 & $[12.0,\ 12.2]$  \\
RPCDM & all & $3$ &  0.4 & $[11.8,\ 11.9]$  \\
\hline
$\Lambda$CDM & filament & $3$ & 0 & $[12.3,\ 12.4]$   \\
$w$CDM & filament & $3$ &  0 & $[12.4,\ 12.5]$  \\
RPCDM & filament & $3$ &  0 & $[12.1,\ 12.2]$  \\
\hline
$\Lambda$CDM & sheet & $3$ & 0 & $[11.7,\ 12.1]$   \\
$w$CDM & sheet & $3$ &  0 & $[11.6,\ 12.2]$  \\
RPCDM & sheet & $3$ &  0 & $[11.5,\ 11.8]$  \\
\enddata
\label{tab:mflip}
\end{deluxetable}

\end{document}